\documentclass[12pt,letterpaper]{article}
\usepackage{amsmath,amssymb,pgf,pgfarrows,pgfnodes,float,appendix, hyperref}
\usepackage{graphicx}
\usepackage{subfigure}
\usepackage[margin=0.9in]{geometry}

\newcommand{\be}{\begin{equation}}
\newcommand{\ee}{\end{equation}}
\newcommand{\bea}{\begin{eqnarray}}
\newcommand{\eea}{\end{eqnarray}}

\title{{\rm\footnotesize \qquad \qquad \qquad \qquad \qquad \ \qquad \qquad \qquad \ \ \ \ \ \                  UTTG-09-15\ TCC-002-15     RUNHETC-2015-7     
SCIPP 15/06}\vskip.5in     CP Violation and Baryogenesis in the Presence of Black Holes}
\author{Tom Banks\\
Department of Physics and SCIPP\\
University of California, Santa Cruz, CA 95064\\
{\it and}\\
Department of Physics and NHETC\\
Rutgers University, Piscataway, NJ 08854\\
E-mail: \href{mailto:banks@scipp.ucsc.edu}{banks@scipp.ucsc.edu}
\\
\\
W. Fischler\\
Department of Physics and Texas Cosmology Center\\
University of Texas, Austin, TX 78712\\
E-mail: \href{mailto:fischler@physics.utexas.edu}{fischler@physics.utexas.edu}}
\date{}
\begin{document}
\maketitle

\begin{abstract} In a recent paper\cite{fk} Kundu and one of the present authors showed that there were transient but observable CP violating effects in the decay of classical currents on the horizon of a black hole, if the Lagrangian of the Maxwell field contained a CP violating angle $\theta$.  In this paper we demonstrate that a similar effect can be seen in the quantum mechanics of QED: a non-trivial Berry phase in the QED wave function is produced by in-falling electric charges. We also investigate whether CP violation, of this or any other type, might be used to produce the baryon asymmetry of the universe, in models where primordial black hole decay contributes to the matter content of the present universe. This can happen both in a variety of hybrid inflation models, and in the Holographic Space-time (HST) model of inflation\cite{holoinflation3}.  

\end{abstract}

\section{Introduction}

The CP violating term \begin{equation} \frac{\theta_{QED}}{32 e^2 \pi^2} \int F \wedge F , \end{equation} is a total derivative, $ \int F\wedge F = \int_{\cal B} A\wedge F , $ where ${\cal B}$ is the boundary of space-time.  It is widely believed that this term has no effect on physics, except in the presence of magnetic monopoles\cite{Witten}.  However in \cite{fk}, Kundu and WF showed that the $\theta_{QED}$ term induces a vorticity in the transient flow of classical current induced on a black hole horizon by a charge falling through it.There is no real contradiction with conventional wisdom, because it is obvious that for supported observers outside the black hole, the horizon is a non-trivial boundary. This is the basis of the Membrane Paradigm\cite{membraneparadigm}. Indeed, the vorticity observed by \cite{fk} was proportional to the Chern-Simons invariant on the horizon. 

Our aim in the first part of this paper is to show that similar effects can be seen in the quantum mechanics of electrodynamics in the presence of horizons.  As first noted by Jackiw\cite{jackiw}, in temporal gauge the effect of the $\theta_{QED}$ term is to induce a Berry phase\cite{berry} into the wave functional.  The phase is proportional to the Chern-Simons term and can have observable effects if there is a non-trivial closed loop in the configuration space of fields.  We will show that the infall of charged particles produces such a closed loop.  

One may ask whether these phenomena might have some effect in cosmology.  More generally, one expects CP violating physics to be important near the horizon of a black hole, if there are CP violating  terms of {\it any} dimension in the effective Lagrangian.  Since black hole decay violates baryon number, it is interesting to ask if a population of primordial black holes could contribute to the baryon asymmetry of the universe. In the late 70's Barrow\cite{barrow} started to investigate baryogenesis 
from primordial black holes.  Hook\cite{hook} has recently explored this possibility.  He works in the context of hybrid inflation models, some of which produce a primordial black hole population and uses the mechanism of {\it spontaneous baryogenesis}\cite{kaplan}. 

The point is that, to leading order, the emission of Hawking radiation is a thermal process, and CPT prevents thermal emission of a baryon excess. Hook uses the violation of CPT by the expansion of the universe to overcome this objection.  In fact however, by the very nature of the fact that it is a {\it decay}, Hawking radiation is not exactly thermal.  The Hawking temperature changes with time and the emission spectrum is not exactly thermal. We should expect a baryon excess to be generated at time $t$, proportional to some positive power of $|{dT/dt}|$.  We will argue, in a very general way that the power is $1$.  

The implications of {\it Hawking induced baryogenesis} for hybrid inflation models depends on details of the model.  The primordial black hole density and the reheat temperature of the universe depend on different regions in the inflaton potential.  The baryon asymmetry produced in black hole decay will also be model dependent.  Thus one must explore a range of models to find whether there is a ``plausible" regime where the correct baryon asymmetry is produced without violating other constraints.

By contrast, in the HST model of the early universe, a primordial population of black holes is produced automatically, and dominates the energy density of the universe until the black holes decay.  The radiation dominated Big Bang is produced by black hole decay, and the reheat temperature of the universe is related to the size of primordial curvature fluctuations by a formula involving a small number of universal parameters\cite{holoinflation3}.  The amount of baryon asymmetry generated depends on one new effective parameter, the ratio of CP violating to CP conserving matrix elements of the Hamiltonian responsible for black hole decay.  We show that if this ratio is $o(1)$ we can reproduce the observed asymmetry within the uncertainties of the calculation\footnote{The back of the envelope estimates of inflationary parameters given in \cite{holoinflation3} produce an asymmetry of order $10^{-9}$, but the uncertainties in these estimates are large.}.  Given these estimates, it is unlikely that the CP violation induced by $\theta_{QED}$, which are of order $\frac{\alpha_{em}}{\pi}$ could be responsible for the baryon excess, but primordial black holes in the early universe communicate with much more general sources of CP violation.

\section{The Berry Phase}

The addition of the total derivative term to the action does not change the equations of motion, but it changes the canonical momenta as a function of the time derivatives of the vector potential.  In any coordinate system it shifts the definition of the canonical momentum by
\begin{equation} \Pi_i \rightarrow \Pi_i - \theta_{QED} B_i  . \end{equation} 
When we choose temporal gauge, in any coordinate system, this shift amounts to \begin{equation} \frac{\delta}{i\delta A_i (x)} \rightarrow \frac{\delta}{i\delta A_i (x)} - \theta_{QED} B_i  . \end{equation}   We can compensate for this shift in the Schrodinger equation by multiplying the wave functional by a phase
\begin{equation}  \Psi [A_i ] \rightarrow e^{ i \theta_{QED} \int A_i B_i } \Psi [ A_i ] . \end{equation} The phase is the Chern-Simons action of the three dimensional gauge fields.
As pointed out by Berry\cite{berry} such an overall phase can lead to observable effects if there is a closed loop in the configuration space such that the phase does not come back to itself.  The fact that $\theta_{QED}$ produces a Berry phase in the QED wave function, was pointed out by Jackiw\cite{jackiw}.

Now let us consider a horizon and study the metric in the near horizon limit.
For Schwarzschild black holes, and the horizon of de Sitter space, the near horizon geometry approaches that of Rindler space\footnote{For genuine Rindler space, the horizon radius $R$ in the formula above goes to infinity and the transverse geometry is flat.}.
\begin{equation} ds^2 \rightarrow - r dt^2 + \frac{dr^2}{r} + R^2 d\Omega^2  . \end{equation}
In the above formula, we restrict attention to $r > 0$.  This coordinate system is appropriate to an accelerated detector, which never falls through the horizon.  For a supported detector, the horizon is a genuine boundary of space-time and the Membrane Paradigm\cite{membraneparadigm} shows us that treating it as such captures all of the physics of the horizon, as seen by the detector.  In fact, in the Membrane paradigm, the boundary is taken to be the {\it stretched horizon}, a timelike hyperboloid whose space-like distance to the horizon is a few Planck lengths.

Consider a process in which a charged particle falls through the stretched horizon. The electric field 
\begin{equation} E_r = \partial_t A_r  , \end{equation} is, in this gauge, entirely due to a time dependent vector potential $A_r$.  We can always write $A_r = \partial_r W$, where $W$ is the open radial Wilson line running from infinity to $r$.  $W$ will be time dependent when we have an infalling charged particle.  The initial value of $W$ near $r = 0$ vanishes.

Now, for a very large black hole, consider a time $\ll R {\rm ln} (RM_P)$ after the particle has passed through the {\it stretched horizon}\cite{membraneparadigm}. $W$ will have changed by a finite amount.  In addition, because current has passed through the stretched horizon, the magnetic field $F_{\theta\phi}$ is non-zero there. Note that so far we have not made a closed loop in configuration space.
The phenomena noted in \cite{fk}, are therefore not connected to the Berry phase.  Indeed, the $\theta_{QED}$ parameter multiplies a total time derivative in the action plus a total spatial divergence.   The Hall current of \cite{fk} is related to the integral of the spatial divergence in the bulk (in the temporal gauge this appears as a modification of Gauss' law), while the Berry phase is related to the time derivative in the Lagrangian.  If we now take $t$ larger than the scrambling time, so that $F_{\theta\phi}$ vanishes on the stretched horizon, we have performed a closed loop in configuration space.  $A_r$ has changed from zero to a pure gauge with gauge function equal to the asymptotic value of the Wilson line on the stretched horizon.   The other fields vanish and so does the Berry phase.  

Now consider the same system in the presence of an external magnetic field, constant in time, with non-zero components $F_{\theta\phi}$.  This field might be generated by a neutron star orbiting the black hole, or simply a stationary configuration of plasma.   In addition, rather than dropping one particle into the hole, we start with no charges, create a pair, separate them and drop each into the horizon.  The final configuration has only the static magnetic field and 
$A_r = \partial_r (W_{12}).$  $W_{12}$ is a Wilson line which goes from one point on the stretched horizon to another.  More precisely, the $r$ dependent $W_{12}$ in the vicinity of the stretched horizon is the Wilson line on a shifted stretched horizon, a distance $r$ from the actual horizon.  From the point of view of the supported observer, the charged particles take an infinite time to reach the horizon.  They reach the stretched horizon in a finite time.  Using Stokes' theorem, we can write the Wilson line describing the creation of the charged particle pair and their subsequent fall through the stretched horizon in terms of a Wilson line connecting them on a fixed surface of constant\footnote{We may pick up contributions far from the horizon, corresponding to the flux of ambient electromagnetic fields through the closed loop, but these are irrelevant to the present discussion.} $r$.  The vector potential $A_r$ on the stretched horizon $r = x L_P$ is the difference of two such Wilson lines.  After the particles have fallen through the stretched horizon it becomes time independent, and therefore pure gauge.  The presence of a boundary at $r = x L_P$ and the assumption that the field $F_{\theta\phi}$ is non-zero on the stretched horizon both before and after the infall tells us that we have performed a closed loop in the configuration space of fields outside the stretched horizon, with a non-trivial Berry phase. 

The Berry phase has evolved from $0$ to \begin{equation} \theta_{QED} \int \partial_r W_{12} F_{\theta\phi} = \theta_{QED} \int d\theta ^ d\phi W_{12} F_{\theta\phi}  . \end{equation}  Here we've used the fact that there is no monopole source for the magnetic field. Generically this will be non-vanishing because the two points on the horizon can be chosen arbitrarily, as can the external field.  
The way in which the horizon provides a non-trivial two sphere in space-time, such that the change in Chern-Simons invariant over this sphere is physical, is reminiscent of the way that black hole horizons force us to include monopole $U(1)$ bundles in the low energy effective field theory of Maxwell's equations\cite{tbseiberg}.

We have thus demonstrated that quantum processes in the presence of a horizon have amplitudes which are sensitive to the value of $\theta_{QED}$ and thus violate CP.   These phases are obviously of order $\alpha_{em}$.

\section{Baryogenesis in Hawking Radiation}

At first sight, the idea that we could generate a baryon asymmetry\footnote{Everywhere in what follows, we use $B$ instead of $B - L$ to save writing.  Electroweak $B$ violation implies that the only interesting primordial asymmetry is that in $B - L$. Our arguments apply equally well to this quantum number.} in the decay of a black hole seems wrong even in the presence of CP violation and baryon number violation.  To a good approximation the radiation is thermal, and CPT
guarantees that a thermal distribution has equal probabilities to be a baryon or an anti-baryon. On the other hand, we are talking about a {\it decay}, and decays violate CPT by their very nature.  In a CPT invariant state, every decay is balanced by a recombination event.  Indeed, the quintessential example of a baryogenesis mechanism is the out of equilibrium decay of an unstable particle.
There is indeed no baryogenesis in the unstable equilibrium state (the Hartle-Hawking state) of a black hole in a radiation bath.  However, a decaying black hole is only described approximately by this state.

The consistency between the two contradictory sentences in the previous paragraph lies in the phrase ``to a good approximation".  Black hole decay is thermal only in the approximation that its rate is slow, so that the temperature changes adiabatically.  Thus, we may contemplate producing a baryon asymmetry in black hole decay, at a rate which is proportional to some positive power of the rate of change of temperature $dT/dt$.  The question we need to answer is, ``What is the power?" 

The decay of black holes has a number of features, which distinguish it radically from the decay of lepto-quarks in old fashioned grand unified models of baryogenesis.  There is no sense in which the early universe is an equal mixture of black holes and ``anti-black holes"\footnote{Here we consider neutral black holes.  We'll return to the case of charged black holes in the next section.}.   We will first discuss the decay of a single black hole and then argue that we can simply sum this result over the gas of black holes produced in the early universe by your favorite mechanism.  The eigenstates of the black hole cannot be assigned a baryon number, because baryon number has no reason to be even approximately conserved by the quantum gravity dynamics that determines the black hole spectrum. The same is likely to be true for CP\footnote{Even in standard model processes, the improbability of CP violation is not a consequence of the intrinsic smallness of CP violating matrix elements, but rather of the hierarchy of masses and mixings of the quarks.}.  Nonetheless, CPT guarantees that, in the thermal approximation, the expected value of baryon excess vanishes.  There are fluctuation corrections to this, but these must be averaged over all the black holes in the universe, and are typically very tiny\footnote{We suggested that these corrections might be sufficiently large in \cite{holoinflation3} but upon more detailed study we found them to be negligible.}.  

The baryon excess should have a power series expansion in the black hole decay rate $\frac{d {\rm ln M} }{dt} $, and CPT indicates that only odd powers contribute\footnote{In the completely unrealistic scenario in which CP violation is small {\it and} black hole decay vanishes in the limit of CP conservation, we would get a square root dependence.}.  The dominant term will be the first power and there is no reason to believe it should be zero.  The result for the instantaneous change in baryon number in a black hole decay should be proportional to the black hole area, the CP violating parameter, and $dT/dt$ the intensive signal of the fact that the decay is out of equilibrium.
We get
\begin{equation} \frac{dB}{dt} = - \epsilon_{CP} M^2 M^{-2} \frac{d(M/M_P) }{dt} , \end{equation} where $\epsilon_{CP}$ is the strength of CP violating processes.  Thus, the amount of baryon number produced in the decay of a single black hole of large mass is 
\begin{equation} \Delta B = - \epsilon_{CP} (M/M_P) . \end{equation}
This has a universal sign for neutral black holes, and as long as the black hole gas is sufficiently dilute that Hawking particles emitted by one black hole have a negligible probability to be absorbed by another, it will simply be additive over the whole universe.
The initially produced baryon density is thus
\begin{equation} \Delta b = \epsilon_{CP} n_{BH}  (M/M_P) . \end{equation} $n_{BH}$ is the initial number density of black holes, which is less than $M^{-3} M_P^6$.
The black holes will evolve as a non-relativistic gas until they decay at a time
$\tau \sim \frac{M^3}{M_P^4} $.  The inflationary era, whether hybrid or HST, which produced the black holes, will also produce a radiation gas at a temperature $T_{RH}$ .  For the HST model, this primordial $T_{RH} = 0$.

At time $\tau$ the black hole energy density is $M n_{BH} a^{-3} (\tau )$ , while the radiation energy density is $g T_{RH}^4 a^{-4} (\tau )$, where $g$ is the effective number of massless particle states into which the black hole decays.  At this point the black holes have all decayed, so we have a radiation gas at temperature
\begin{equation} g T^4 = g T_{RH}^4 a^{-4} (\tau ) + M n_{BH} a^{-3} (\tau ) . \end{equation}
The baryon to entropy ratio is
\begin{equation} \frac{\Delta b}{\sigma } = (g T^3 )^{-1} n_{BH} \epsilon_{CP} (M/M_P) a^{-3} (\tau ) . \end{equation}

There are two simple limits in which we can evaluate these formulae.  In the first, the black hole energy density dominates the primordial radiation at $\tau$ and we have 
\begin{equation} \frac{\Delta b}{\sigma } = (g )^{-1/4}M^{-3/4} n_{BH}^{1/4} \epsilon_{CP}  (M/M_P) a^{-3/4} (\tau ) . \end{equation}  If, as in the HST model, the black hole energy dominated the radiation initially, then $a^{-3/4} (\tau ) = (\frac{3}{2} \sqrt{\frac{8\pi M n_{BH}}{3 M_P^2}} \tau + 1)^{- 1/2} $.  In this limit, the initial black hole number density $n_{BH}$ drops out of the formula for the baryon asymmetry.  Up to ``order $1$" numerical factors, we get
\begin{equation} \frac{\Delta b}{\sigma } = (g )^{1/4}(\frac{M}{M_P})^{-3/2} \epsilon_{CP}  . \end{equation}  In the HST model, $M$ is about $10^6 M_P$.  Taking $g \sim 10^3$ we get
\begin{equation} \frac{\Delta b}{\sigma } \sim 6 \times (10)^{-9} \epsilon_{CP}  . \end{equation} 
For $\epsilon_{CP}$ of order $1$ this overshoots the observed value a bit, while for $\epsilon_{CP} \sim \frac{\alpha_{QED}}{\pi}$ as one would expect from an order one $\theta_{QED}$ it undershoots by a factor $\sim 10 - 100$.  Given the crudeness of our estimates, this has to be considered a success, though the attribution of the entire baryon asymmetry to $\theta_{QED}$ in this model, seems a bit problematic.   The formulae above are applicable to primordial black holes from hybrid inflation as well, as long as the energy density in black holes dominates that in radiation until the time of black hole decay.  The value of the average black hole mass $M$ can take on a variety of values in different hybrid inflation models, and we have not made a survey to determine which of these models could account for the baryon asymmetry by our mechanism.

We note that Hook's\cite{hook} formulae for the baryon yield differ dramatically from our own, particularly in their dependence on the reheat temperature.  This is easy to understand.  For Hook, the thing driving the departure from equilibrium is the expansion of the universe, and a low reheat temperature means that the expansion is slow when most of the black holes decay.  By contrast, in our model, departure from equilibrium is independent of the expansion rate, and the reheat temperature appears only in the denominator of the baryon to entropy ratio.  Hook's baryon yields scale with a positive power of $T_{RH}$ while ours scale as $T_{RH}^{-4}$ .  

Another simple limit to study is one in which the black hole energy density is always {\it smaller} than that of the radiation gas.  This will not occur in HST, but could occur in some hybrid inflation models.  In this case we have

\begin{equation} g T^4 = g T_{RH}^4 a^{-4} (\tau ) . \end{equation}
The baryon to entropy ratio is
\begin{equation} \frac{\Delta b}{\sigma } = (g T_{RH}^3 )^{-1} n_{BH} \epsilon_{CP} (M/M_P) a (\tau ) . \end{equation} In addition
\begin{equation} a(\tau ) =  (\frac{32\pi}{3})^{1/4}\frac{T_{RH}}{M_P}(M_P\tau )^{1/2} = (\frac{M}{M_P})^{3/2}(\frac{32\pi}{3})^{1/4}\frac{T_{RH}}{M_P} g^{- 1/2} . \end{equation}
The final result is 
\begin{equation} \frac{\Delta b}{\sigma } = (\frac{32\pi}{3})^{1/4} g^{- 3/2} (\frac{M}{M_P})^{5/2} \frac{n_{BH}}{T_{RH}^3} \epsilon_{CP} . \end{equation}
Different hybrid inflation models will have different values of $M, g$ and $T_{RH}$ and we are not sufficiently familiar with the hybrid inflation literature on primordial black hole production to do a survey.  However, the constraints that the black hole energy density never dominate the radiation density, combined with the inequality $M > M_P > T_{RH}$ suggest that the factor 
$$ \frac{n_{BH}}{T_{RH}^3} $$ will be very small in the regime where the above estimate applies.  It seems unlikely that models satisfying these constraints could give a large enough asymmetry.   We have not studied the intermediate case, where black holes are initially subdominant, but come to dominate the energy density before they decay.

\section{Primordial Charged Black Holes}

CP violating effects of a non-vanishing $\theta_{QED}$ are enhanced by the presence of magnetically charged black holes, and such black holes also catalyze baryon number violation via the Callan-Rubakov effect.  In an ancient iteration of the HST cosmology, we claimed that the universe would be populated by a very dilute gas of very heavy {\it monop-holes } with large magnetic charge and speculated that they could produce the baryon asymmetry.

Our current understanding suggests that those claims were incorrect.  They were based on a too-literal reading of the phrase {\it dense black hole fluid}, which we used to analyze the properties of the $p = \rho$ phase of cosmic evolution.  We also mixed up effective field theory notions with our new holographic models in a way that makes no sense.   The proper way to analyze the probability of forming a magnetic charge, whether for the horizon filling black hole of the $p = \rho$ era, or the individual post-inflationary black holes of \cite{holoinflation3} is to use the black hole entropy formula but only after the $p=\rho$ and inflationary eras have ended.  In the model of \cite{holoinflation3}, the universe has no localized black holes, to which this formula could apply, until the end of inflation.

The entropy formula gives the probability that a black hole of fixed mass has integer electric and magnetic charges $(n_e, n_m)$ as
\begin{equation} P[(n_e, n_m)] = e^{ -\frac{1}{4} \alpha [\ (n_e + \frac{\theta}{2\pi} n_m)^2 + \frac{1}{16\pi^2 \alpha^2} n_m^2]} . \end{equation}  Even if we use the value of the fine structure constant at the unification scale, then as long as perturbative unification is valid, the probability of magnetic charges is very small and it would seem that only the lowest value $n_m = 1$ could be of any significance. Of course, estimating the probability from the black hole entropy is really only valid for macroscopic amounts of charge, so what we can really conclude is that only $o(1)$ values of the monopole charge, where the black holes are far from extremal, will be present with $o(1)$ probability.  

On the other hand, the decay of such black holes will, if $\theta_{QED}$ is $o(1)$ violate both baryon number and CP by amounts of order $1$ per unit time.
Recall\cite{callanrubakov} that a Planck mass monopole black hole will create a distorted region of QCD vacuum around it whose size is much larger than the Schwarzschild radius.  Particles created in the decay of the more numerous neutral black holes will have interaction cross sections at the QCD scale with this cloud and be sucked into the region where we can see explicit CP violation from the monopole's electric field, as well as explicit baryon violation at the black hole horizon.

The decay process of a monop-hole can be modeled like that of an X boson in grand unified models, except that since the decay amplitudes are not perturbative, there is no analog of the loop suppression of CP violation in those Feynman diagram calculations\cite{weinbergetal}.  We proceed from a thermal soup of magnetically neutral and magnetically charged black holes.  Let $f_m$ denote the fraction of holes with non-zero positive magnetic charge.  There's an equal number of negative charges.   The basic process that leads to baryogenesis is the transition
\begin{equation} (M , n_m) \rightarrow (M - \Delta M , n_m) + \Delta B , \end{equation} where $\delta B$ is a collection of particles with total energy $\Delta M \ll M$ in the monop-hole rest frame, and baryon number $\Delta B$ .  Order $1$ CP violation means that the process
\begin{equation} (M , \bar{n}_m) \rightarrow (M - \Delta M , \bar{n}_m) - \Delta B , \end{equation} has a different probability.  These probabilities are not computed in a perturbative loop expansion and they are of order $1$. As in $X$ boson decay, the very fact that the monop-holes are decaying means that we are out of thermal equilibrium.  The black hole gas is dilute, and inverse processes in which particles emitted by one black hole are absorbed by another, are much more rare than the decay processes.  In the approximation that we neglect them, the Callan-Rubakov effect seems unimportant.

Also, the fact that the black holes are charged means that we do not require a factor of $dT/dt$ in the decay rate, in order to account for the asymmetry. We can think of the black hole charge as a chemical potential, shifting the equilibrium of black hole emission to one that prefers particles over anti-particles.   The rate of baryon violation in single monop-hole decay is thus

\begin{equation} \frac{dB}{dt} \sim  (M/M_P)^2 \Delta P , \end{equation} where $\Delta \frac{dP}{dt}$ is the difference in probabilities per unit time for the two processes above. This difference should be $o(1)$ if $\theta_{QED}$ is $o(1)$.
This should be multiplied by $f_m n_{BH}$ to calculate the total baryon asymmetry of the universe produced by primordial monop-hole decay.

We thus have \begin{equation} \frac{\Delta B_{neutral}}{\Delta B_{monop-hole}} \sim \frac{\epsilon_{CP\ -\ neutral} }{\epsilon_{CP\ -\ monop-hole} } \frac{M_P}{f_m M} . \end{equation}   If $f_m > M_P / M $, then baryogenesis via monop-hole decay dominates over that from the much more numerous neutral black holes.  This implies $M/M_P > 10^{11} $, which is not compatible with the HST model, where $M/M_P \sim 10^6$. This scenario is perilously close to violating the Parker bound on the monopole density. We have 
$\sim (M_P / M)^2$ monop-holes per baryon.  The relevant form of the Parker bound, for the extremal monop-hole mass $\sim M_P$\footnote{This is {\it not} the mass $M$ of the non-extremal hole before Hawking evaporation has taken place.}, and a presumed velocity $v$ relative to the earth of $\sim 3 \times 10^{-3} c$\cite{parkerturner} is 
\begin{equation} n_{monop-hole} < 10^{-20} (cm)^{-3} \frac{3 \times 10^{-3} c}{v} . \end{equation} Given the observed baryon density
\begin{equation} n_{b} = 10^{-5} (cm)^{-3} , \end{equation} we get a bound
\begin{equation} (M/ M_P)^2  > 10^{15} \frac{v}{3 \times 10^{-3} c} . \end{equation}
The condition that most baryons be produced in monop-hole, rather than neutral black hole decay is $M/M_P > f_m^{-1} \sim 10^{11}$, which is much stronger than the observational bound.  The bounds from monopole catalyzed baryon decay in stars and Jovian planets give bounds on the monopole density $3$ to $14$ orders of magnitude stronger than the Parker bound\cite{pdg}, while the theoretical constraints of our scenario only give us about $7$ orders of magnitude leeway.  Depending on whether one believes the strongest astrophysical constraints, one might be forced to conclude that $M/M_P > 10^{15}$.  The black hole life-time for such black holes is uncomfortably long $ \sim 10 $ seconds, and the reheat temperature from black hole decay is too low to accommodate nucleosynthesis.  It is very unlikely that a model with radiation domination at a temperature $\sim 10$ MeV, could ever be constructed in the presence of such massive monop-holes.  We conclude that the origin of the baryon asymmetry is unlikely to be the decay of primordial magnetically charged black holes.

The probability of finding electrically charged black holes with charge $< \frac{1}{\alpha_{unification}}$ is of order $1$, according to the black hole entropy formula.  These will not exhibit large CP violation merely in response to an order $1$ value of $\theta_{QED}$, but their decays certainly violate baryon number and would also violate CP since there is evidence for $o(1)$ CP violation in the CKM matrix.  Thus, initial decays of the charged black hole will produce an asymmetry.

However, there are two crucial differences when compared with the magnetically charged holes.  First of all, the chemical potential due to the charge, which can be thought of as the driver of CPT violation, is small because the black hole is so far from its extremal limit\footnote{We neglected to include a similar suppression in the calculation of monop-hole decays, because it is somewhat less severe.  At any rate, it only worsens the phenomenological problems of that scenario.}.  Secondly, because electrons are light compared to the initial Hawking temperature, the hole will discharge itself.  In the small charge regime this will happen exponentially rapidly\cite{hook,gibbons}. Thus, the baryon asymmetry produced by this mechanism will be of order zero in $M/M_P$ and is less important than the charge independent asymmetry produced by the shifts in the black hole equilibrium state.  

\section{Conclusions}

We've examined two different questions in this paper, because we initially thought there could be a connection between them.  The first was evidence for
non-trivial effects of the CP violating angle $\theta_{QED}$, in the presence of black hole horizons, even when no monopoles are present.  We found a non-trivial Berry phase in processes where neutral systems with separated charges were dropped into the black hole. We then attempted to see whether CP violation via $\theta_{QED}$ could be the origin of baryogenesis, in models where primordial black holes are produced in the early universe.  

We found that the answer to the second question was negative but we also discovered that, in the presence of other sources of CP violation, the decay of black holes could produce the asymmetry.  This question was previously studied by Hook\cite{hook}, who considered the case of Hawking radiation in the context of an expanding universe. The expanding universe provides in Hook's proposal, the violation of equilibrium necessary to produce a baryon excess. In the case of HST, the change in temperature in black hole evaporation provides the source of non-equilibrium. We showed that it is quite plausible that a baryon excess of the observed magnitude can be produced in HST and possibly also in a class of hybrid inflation models.

\vskip.3in
\begin{center}
{\bf Acknowledgments }
\end{center}
This work was begun while TB was a guest of the Physics Dept. at Georgia Tech, and completed while he attended the workshop on Quantum Gravity Foundations at KITP-UCSB. He thanks both of those institutions for their hospitality.
 The work of T.B. was supported in part by the Department of Energy.   The work of W.F. was supported in part by the TCC and by the NSF under Grant PHY-0969020

\end{document}